# On-Chip Hotplate for Temperature Control of CMOS SAW Resonators

A. N. Nordin[1], I. Voiculescu[2] and M. Zaghloul[1]

[1] Department of Electrical and Computer Engineering, George Washington University, Washington DC, USA
[2] Department of Mechanical Engineering, City College of New York, New York, USA

*Abstract*-Due to the sensitivity of the piezoelectric layer in surface acoustic wave (SAW) resonators to temperature, a method of achieving device stability as a function of temperature is required. This work presents the design, modeling and characterization of integrated dual-serpentine polysilicon resistors as a method of temperature control for CMOS SAW resonators. The design employs the oven control temperature stabilization scheme where the device's temperature is elevated to higher than $T_{max}$ to maintain constant device temperature. The efficiency of the polysilicon resistor as a heating element was verified through a 1-D partial differential equation model, 3-D CoventorWare® finite element simulations and measurements using Compix® thermal camera. To verify that the on-chip hotplate is effective as a temperature control method, both DC and RF measurements of the heater together with the resonator were conducted. Experimental results have indicated that the TCF of the CMOS SAW resonator of -97.2 ppm/°C has been reduced to -23.19 ppm/°C when heated to 56°C.

## I. INTRODUCTION

The common usage of SAW resonators with RF integrated circuits (IC) has motivated its fabrication using standard IC fabrication processes. Fully integrated CMOS SAW resonators with frequencies ranging from 500 MHz to 1 GHz have been successfully fabricated [1]. For device stability, a small variation of frequency with temperature (less than 1ppm/°C) is often desired. SAW devices employing ZnO as its piezoelectric material have been reported to have negative temperature coefficient frequency (TCF) [2, 3]. Various methods have been employed to reduce the effect of the negative (TCF) for resonators utilizing ZnO such as adding control-circuitry [4], using quartz substrates [2], and layering positive TCF $SiO_2$ layers [3]. For our device, we have proposed a simple solution of an embedded heater so that additional temperature compensation circuitry and alteration of the device structure or layers would not be required. Temperature compensation is achieved by heating the resonator structure to a temperature above its typical operating temperature causing the device to be independent of the ambient temperature.

## II. DEVICE STRUCTURE AND FABRICATION

Fig. 1 shows the structure of a two-port CMOS SAW resonator that consists of input and output interdigital transducers (IDTs), with a bank of shorted reflectors on each side. When a sinusoidal signal is injected at the input port, acoustic waves propagate in the piezoelectric layer above the IDTs in both directions. The acoustic wave is detected and translated back into an electrical signal at the output port. The reflectors minimize losses by containing the acoustic waves within the cavity, creating standing waves. The device's resonant frequency is determined by the periodic distance (λ) of the IDT. The IDTs, reflectors and ground shield were implemented using standard CMOS layers [1].

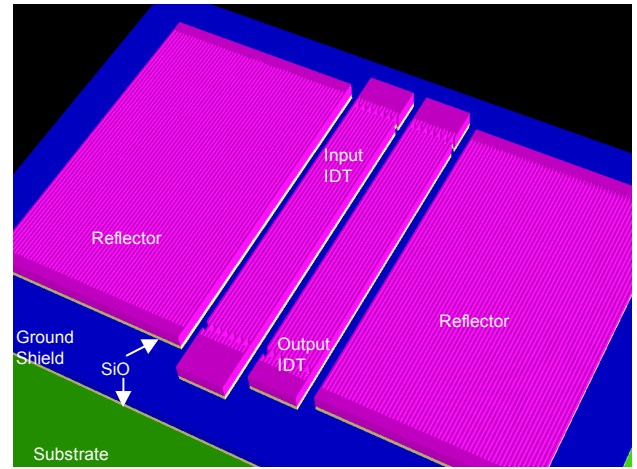

*Fig.1. Implementation of a two-port SAW resonator structure in CMOS. For clarity, the piezoelectric ZnO layer is not shown [1],*

To incorporate a heating element for the CMOS SAW resonator, it is advantageous to utilize material layers inherent in CMOS IC technology. Polysilicon with sheet resistivity of 22.8 Ω/sq. [5], is commonly used as a heating element for CMOS MEMS devices. The choice of polysilicon as the heater layer for our device is ideal since it does not disturb the present device layers. An added advantage is that the ground shield layer now not only provides isolation of the resonator from electromagnetic feedthrough but also acts as an even temperature distributor. This resonator with temperature control elements was fabricated using AMI C5 0.6 μm CMOS process. A series of IC compatible post-processing steps were implemented to integrate the piezoelectric ZnO layer on top of the resonator structure. These process steps were detailed in [1]. The final cross-section of the device is illustrated in Fig. 2.

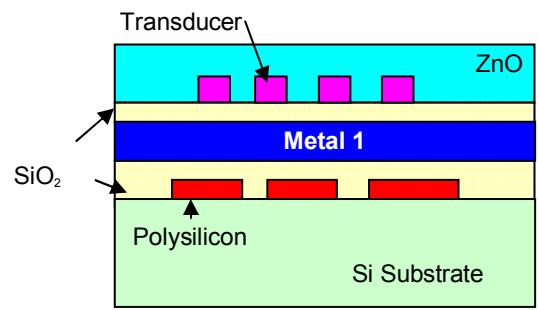

*Fig. 2. Cross-section of CMOS SAW Resonator with embedded heaters*

### III. THEORY AND SIMULATION

To enable both heat generation and *in situ* temperature measurement, two intertwined serpentine heaters have been employed. The schematic for the dual-serpentine polysilicon heaters and its dimensions are shown in Fig. 3 (a). Resistor 1 acts as a heater and Resistor 2 acts a thermistor. The heaters length (*L*) and width (*w*) were obtained from the device's layout. The thickness (*t*) of the polysilicon layer was assumed to be 0.3 μm. The nominal resistances of the heaters were calculated based on the dimensions and the sheet resistivity.

The thermal analysis of a polysilicon heater can be divided into two stages. First, the polysilicon heater is represented as a one-dimensional (1-D) structure shown in Fig. 3 (b) and the heat flow is modeled using partial differential equations. Next, a more complex three-dimensional model of the polysilicon heater is designed in CoventorWare® and analyzed using Finite Element Method (FEM).

Based on Fig. 3 the transient heat flow through a simple heater at steady-state can be simplified into Poisson's partial differential equation as shown in (1)

TABLE 7-1: POLYSILICON HEATER DIMENSIONS AND 1-D STEADY STATE TEMPERATURE

|    | L (μm) | W (μm) | T (μm) | R (kΩ) | Applied Voltage (V) | Temp (K) |
|----|--------|--------|--------|--------|---------------------|----------|
| R1 | 4088   | 12     | 0.3    | 5.05   | 1                   | 708.82   |
| R2 | 3993   | 12     | 0.3    | 4.99   | 1                   | 699.33   |

$$\frac{\partial^2 T}{\partial x^2} = -\frac{Q(x)}{\mathbf{k}} \quad (3.1)$$

The following assumptions were made: the power absorbed and dissipated by the heater is uniform and the ends of the heater have temperatures of $T_{amb}$ = 300K. Based on this, $Q(x)$ in (1) can be replaced with the power generated in the resistor, $P = (I.V/L.w.t)$, where $I$ is the current flowing in the resistor, $V$ is the applied voltage and $L, w, t$ are dimensions of the resistor.

The heat flowing along *x*-axis is assumed to be uniform in the *x-y* plane. If the center of the heater is defined as *x*=0, the heater length at both ends, due to symmetry is +/- *(L/2)*. The power absorbed and dissipated by the heater is also assumed to be uniform. Also due to symmetry, only half of the heater power *P/2* is used for solving (1). This equation can then be integrated twice to obtain the temperature as a function of *x* as shown in (2)

$$T(x) = 300 + \frac{P}{4\mathbf{L} \cdot \mathbf{w} \cdot \mathbf{t} \cdot \mathbf{k}}\left[\left(\frac{\mathbf{L}}{2}\right)^2 - x^2\right] \quad (2)$$

The constants were obtained by applying the following boundary conditions: the ends of the heater are considered to be connected to perfect heat sinks or $T(L/2) = T(-L/2)$ = 300 K. The steady-state temperature of the heaters at midpoint or x=0 was calculated using the (2) and is shown in Table 1.

To improve accuracy, the 1-D model of the serpentine heater was verified using 3-D finite element thermal analysis simulation using CoventorWare®. The polysilicon heaters with dimensions as shown in Table 1 were drawn as a layout. 1 V was applied between the both ends of Resistor 2. Using the FEM thermal analysis, the steady state temperature distribution in the polysilicon layer was obtained as shown in Fig. 4. The maximum temperature occurs at x=0 and has the value of 740 K. This value was close to the 1-D approximation of 699.3 K calculated earlier.

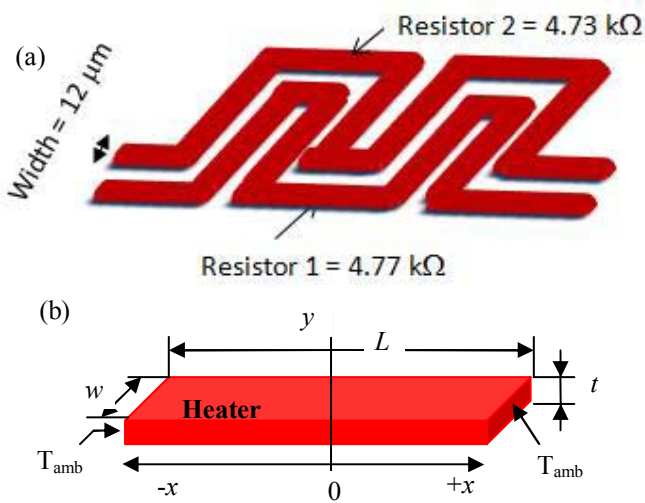

*Fig. 3. (a) Heater geometry and dimensions. (b) Simple model for 1-D heat flow*

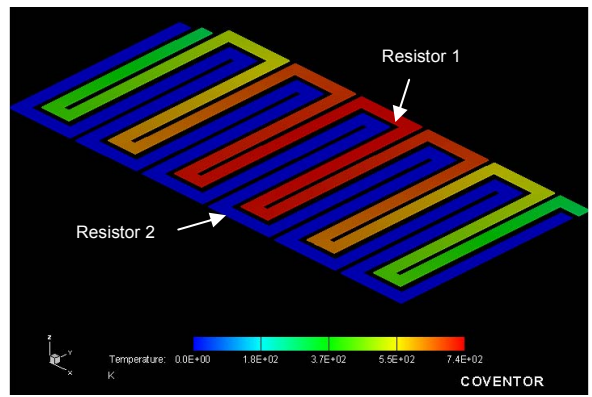

*Fig. 4. Steady-state temperature distribution of polysilicon heater with 1 V applied input.*

TABLE 2: CMOS HEATER LAYERS AND MATERIAL PROPERTIES

| Material | Thickness (μm) | k (pW/μmK) | c (J/kgK) | ρ (kg/m) |
|---|---|---|---|---|
| PolySi | 0.3 | $20 \times 10^6$ | 678 | 2330 |
| SiO$_2$ | 0.6 | 1.2 | 730 | 2270 |
| Al | 1.5 | 190 | 963 | 2699 |

The 1-D model only provides an approximation of the temperature distribution in the polysilicon heaters. In reality due to the CMOS process, the polysilicon heaters are embedded underneath SiO2 and Al layers. To mimic the actual CMOS fabrication process, the device was modeled using a composite of polysilicon-SiO$_2$-Al layers. The thickness of each layer and its material properties used in the simulation was shown in Table 2.

It is expected that the additional layers will introduce heat losses and the temperature increase would not be as high as when only polysilicon layer was used. To compensate this effect, the applied voltage difference to the heater pads was increased to 10 V. The other boundary conditions such as the heater electric pads = 300 K was kept as in previous analyses. Fig. 5 illustrates the steady-state current density flowing through Resistor 2. The model was probed at the center of the device and the current density was found to be 0.14 mA/μm$^2$. The maximum current density in the resistive element was 0.32 mA/μm$^2$.

The effectiveness of the polysilicon resistor as a heating element can be evaluated by observing the temperature distribution in the device. Fig. 6 illustrates the steady-state uniform temperature distribution in the serpentine heaters with the metal conducting metal plate. It can be seen that due to the conducting Al plate, both resistors are heated up although voltage was applied only to Resistor 1. The Al metal sheet also acts as a heat distribution plate, creating an even temperature distribution underneath the entire device. The highest temperature was observed at the center of the device which is also where most of the acoustic waves are concentrated. This indicates that the serpentine heater structure is perfect for our SAW resonator design as it ensures the heat in the micro-oven is focused at the center of the device and avoids unnecessary heating of unimportant parts of the device. The maximum temperature observed at the center of the device was simulated to be 520K.

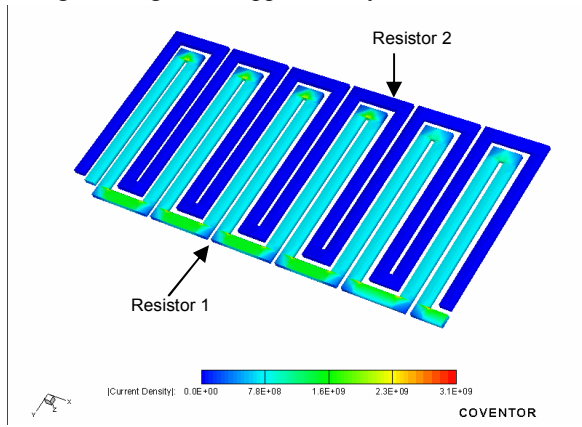

Fig.5. Steady-state current distribution of polysilicon heaters with 25 V applied input.

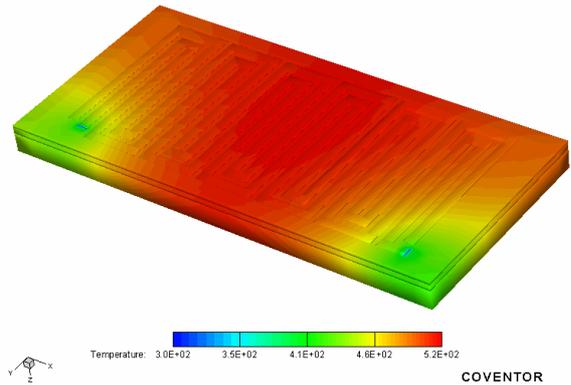

Fig. 7: Steady-state temperature distribution of on ZnO with 25 V applied input at Resistor 1.

An interesting aspect of the usage of two polysilicon resistors is that more efficient heating can be done through applying voltage to both of the resistors simultaneously. Temperature-resistance calibration was performed using temperature measurements with a thermal camera and resistance measurements of Resistor 1. Once calibration has been completed, Resistor 1 no longer needs to work as a thermometer. Both resistors can be used as heating elements. When the two resistors are connected in parallel, the resistance of the circuit reduces and the current increases. The amount of heat flux and temperature increase is almost doubled since both resistors are of similar values. The simulation results with using both elements in parallel indicate temperature increase to 1200K as shown in Fig. 7.

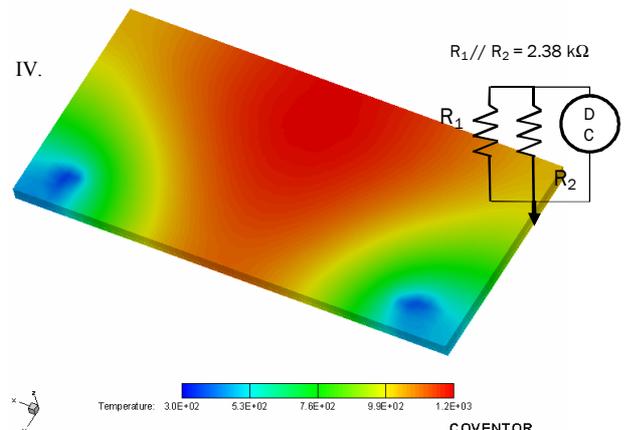

Fig. 7. Steady-state temperature distribution of on ZnO with 25 V applied input both Resistor 1 and Resistor 2. Inset: Parallel arrangement of resistors connected to DC voltage.

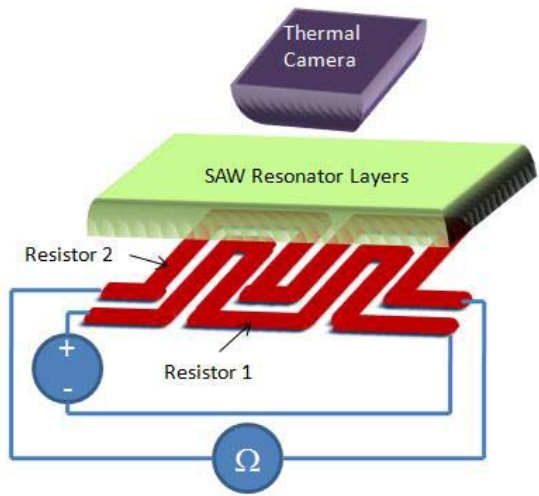

*Fig. 8. Schematic of experimental setup for temperature-resistance calibration. Varying voltage was applied to Resistor 1 and resistance measurements were done on Resistor 2.*

## V. TEMPERATURE CALIBRATION OF THE THERMISTOR

In order for the polysilicon resistor to function as a thermometer and to measure the temperature of the device, the calibration curve which plots the change in resistance at different temperatures has to be obtained. The experimental setup for such purpose is shown in Fig.8.

The rationale of the experiment was that any heat generated by Resistor 1 will be detected and indicated by the change in resistance of Resistor 2. Change in temperature of the device measured using a thermal camera. To facilitate measurements, the CMOS SAW resonators were bonded onto a 28-pin small outline surface mount package as shown in Fig. 9.

The device is heated by applying different voltages to Resistor 1 and the change in temperature of the device was measured using the Compix® PC2000 thermal imaging system which consists of an infra red camera and the WinTES camera control and analysis software. The camera has thermal resolution of 0.1°C/step and can detect feature sizes of 50 μm. This resolution was sufficient for our application since the each heater size was approximately 100 μm x 100 μm. A maximum temperature obtained when the 25 V was connected to resistor 1 was 42.75°C as shown in Fig. 10 (left).

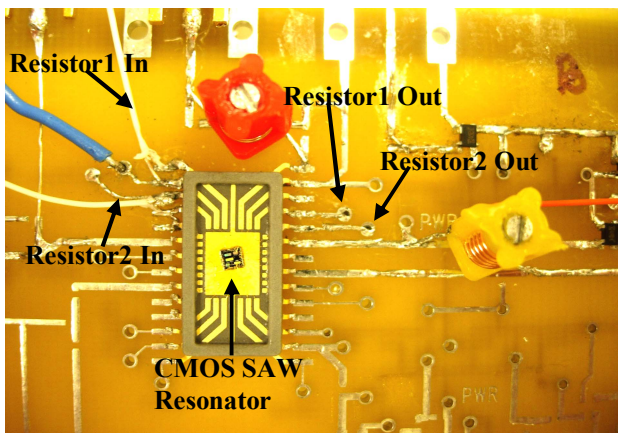

*Fig. 9. Snapshot of packaged CMOS SAW resonator on the printed circuit board.*

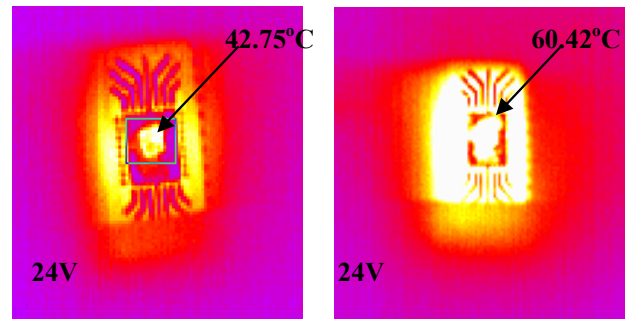

*Fig.10. Thermal camera image of device heated using single resistor (left) and double resistors (right). The applied voltage was 24 V and the corresponding measured temperature of the CMOS SAW Resonator was indicated on the image. Images obtained using the Compix PC2000 thermal imaging system.*

Based on the measured temperature results obtained from the thermal camera images, a 3D plot of the varying voltage versus the measured temperature and the measured resistance can be done and was shown in Fig. 11. The axes of the graph indicate the applied voltage to Resistor 1, measured resistance in Resistor 2 and the device's temperature measured using the thermal camera respectively.

The equation describing the relationship between the applied voltage and corresponding temperature was extrapolated from the graph as (3) where T is measured temperature in $^0$C and V is the applied voltage to Resistor 1. It can be seen that (3) is a third order polynomial.

$$T = 0.0015V^3 - 0.0278V^2 + 0.4614V + 26.594 \qquad (3)$$

Similarly, the expression to describe the function of

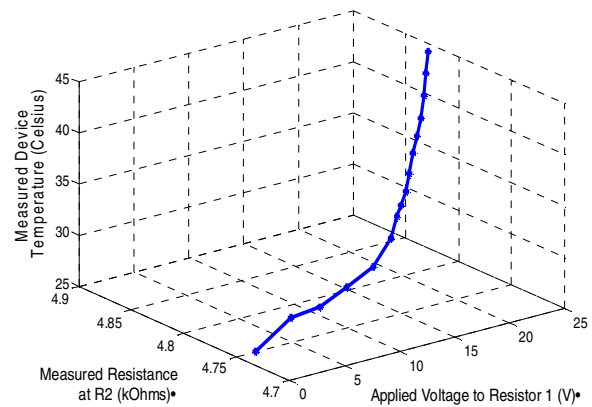

*Fig. 11. Three dimensional plot of applied voltage to Resistor 1 versus measured resistance at Resistor 2 and the device's temperature measured using the thermal camera.*

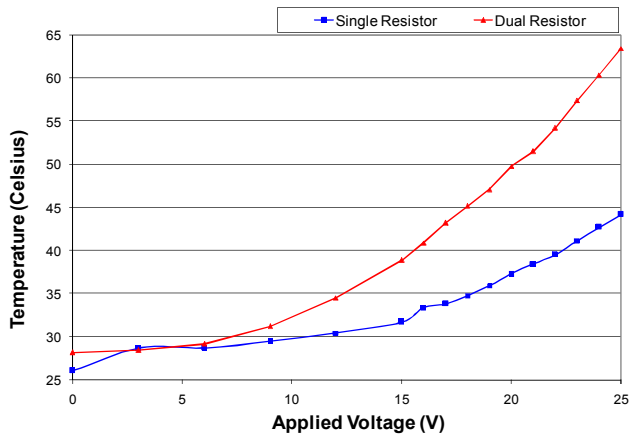

*Fig.12. Measurements of device temperature versus applied voltage to both single resistor and double resistors.*

Resistor 2 as a thermistor was extrapolated from Fig. 11. The linear relationship between the measured resistance in Resistor 2 and the device's temperature is shown as (4) where T is the measured temperature and R is the measured resistance in Resistor 2. Later, during the usage of the CMOS SAW resonator in the field, a thermal camera would no longer be needed to determine the temperature of the chip. Fairly accurate deduction of the device's temperature can be extrapolated from this graph by measuring the resistance in Resistor 2 or can be deduced by the applied voltage at Resistor 1.

$$T = 167.68R - 766.04 \quad (4)$$

When connected in parallel to the DC supply, more efficient heating occurs. The current is now almost doubled and correspondingly the device achieves a higher temperature of $63.52^\circ C$ compared to $44.21^\circ C$ obtained from heating a single resistor. The thermal camera image at the maximum temperature obtained using the parallel arrangement is shown in Fig, 10 (right).

As comparison, the plot of the device's temperature versus applied voltage using a single resistor and both resistors is shown in Fig. 12. The quadratic relationship between the applied voltage and temperature was extrapolated based on Fig. 12 and is shown in (5) where $T_2$ is the device's temperature when two resistors are heated and $V_2$ is the applied voltage to both resistors.

$$T_2 = 0.0679V_2^2 - 0.3046V_2 + 28.455 \quad (5)$$

## VI. TEMPERATURE CONTROL OF THE CMOS SAW RESONATOR

The previous experiments proved that the on-chip polysilicon resistors function very well as heaters, capable of increasing the temperature of the device to $65^\circ C$. The next set of experiments was done to evaluate the efficiency of these heaters to act as a micro-hotplate. The objective is to make the device's resonant frequency be independent of the changing temperatures in the environment. The adopted concept is based on the micro-oven concept of heating the device to a temperature than its typical operating temperature. Such method has been demonstrated to be very successful as a temperature compensation method [6-9].

The first experiment will observe the effect of varying temperatures on the uncompensated device's resonant frequency. In the second experiment, the device was heated to $56^\circ C$ using the polysilicon heaters and then subjected to varying environmental temperatures by placing it on a hotplate. It is assumed that the device will not be subjected to ambient temperatures higher than $56^\circ C$ or $T_{max}$. The fluctuations in the resonant frequency of the device in the both experiments were compared to assess the efficiency of the polysilicon heaters as a method of temperature control.

The experimental setup for both experiments was shown in Fig. 13. In the first experiment, the series resonant frequencies of the device were recorded at different ambient temperatures by applying heat through the hotplate. In this experiment, the ambient temperatures were varied between 32°C to 42°C using the external hotplate. It can be seen in Fig. 14 that due to the negative temperature coefficient of frequency (TCF) for ZnO, an increase in temperature leads to a decrease in resonant frequency.

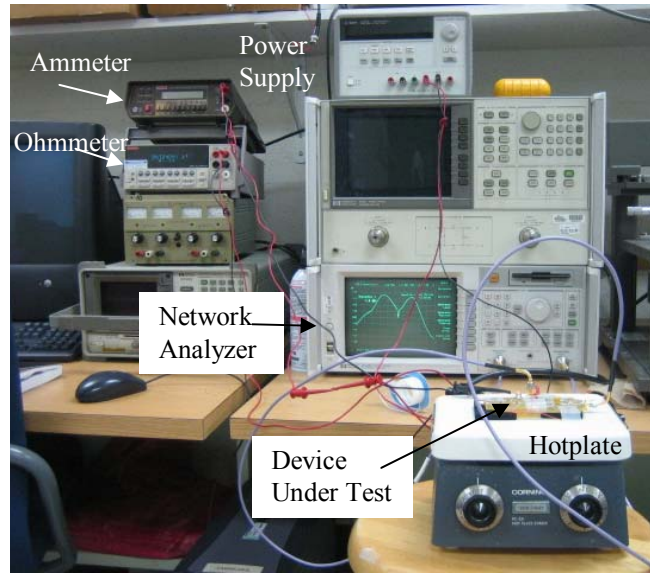

*Fig. 13. Experimental setup for compensated and uncompensated measurements of resonant frequency change with temperature.*

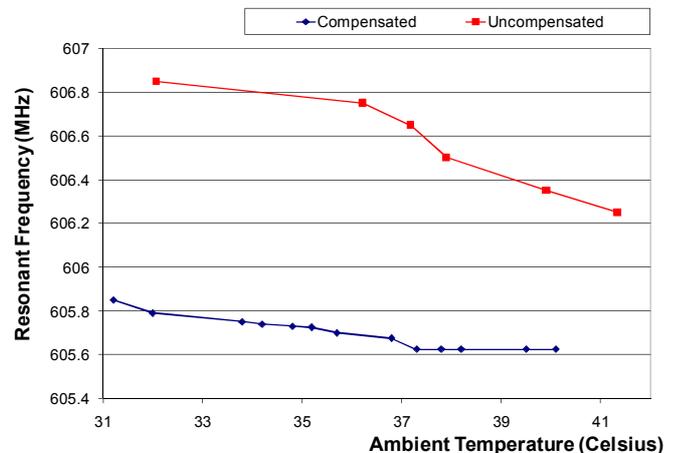

*Fig. 14. Measurement results of resonant frequency change versus change in temperature of the environment.*

The TCF can be calculated using (6) where $\Delta f$ is the change in resonant frequency, $f$ is the resonant frequency at room temperature or 30°C and $\Delta T$ is the change in temperature. The TCF for the uncompensated device was found to be -97.2 ppm/°C

$$TCF = \frac{\Delta f}{f}\left(\frac{1}{\Delta T}\right) ppm/^0C \qquad (6)$$

In the second experiment, the device was heated to 56°C by applying 23V to both heaters. The ambient temperature was varied between 30°C to 40°C. The series resonant frequency of the device was recorded at different temperatures. From Fig. 14, it can be observed that the resonant frequency is less affected by the ambient temperature when the device is heated to $T_{max}$. The compensated TCF was calculated using (6) was found to be -23.19°C. This indicates that the micro-hotplate strategy of stabilizing the resonant frequency of the CMOS SAW resonator is effective in reducing the TCF.

VII. CONCLUSION

A simple method for temperature control using poysilicon micro-hotplate for CMOS SAW resonators was demonstrated. Temperature compensation method for such resonators is important due to the sensitivity of the piezoelectric layer to temperature, causing the resonant frequency to fluctuate. The micro-hotplate elevates the device's temperature to $T_{max}$, causing the resonator to be independent of the varying ambient temperature of less than $T_{max}$. 1-D temperature approximation of the heater was performed to obtain a rough estimate of the temperature increase in the device. This approximation was fine-tuned using 3D FEM simulations to obtain more accurate temperature distribution in the device. Temperature calibration curves for the resistors were attained through using resistive measurements and thermal imaging. Verification of the micro-hotplate's functionality as a method of temperature control was determined experimentally and comparisons between uncompensated TCF and compensated TCF were made. This method of temperature control was successful in reducing the TCF of the device from -97.2 ppm/°C to -23.19 ppm/°C when heated to 56°C.

ACKNOWLEDGMENT

All post-CMOS processing utilized the facilities of the National Nanofabrication Infrastructure Network labs at Howard University, University of California Santa Barbara, University of Minnesota and Georgia Institute of Technology. The authors wish to thank Prof. Gary Harris for the use of the facilities at Howard University and James Griffin for his expertise, assistance and continuous support during fabrication of the resonators. The authors are also grateful for the post-CMOS processing work performed by Dr. Brian Thibeault and Dr. Gregory Book of UCSB and Georgia Tech respectively. The authors are also appreciative for the help extended by Mazdak Taghioskoui and Shumin Zhang during the thermal measurements.